# Normal state properties of high angle grain boundaries in (Y,Ca)Ba$_2$Cu$_3$O$_{7-d}$

S.H. Mennema, J.H.T. Ransley, G. Burnell, J.L. MacManus-Driscoll, E.J. Tarte, M.G. Blamire

Department for Materials Science and Metallurgy, University of Cambridge, Pembroke Street, CB2 3QZ, Cambridge, United Kingdom.

By lithographically fabricating an optimised Wheatstone bridge geometry, we have been able to make accurate measurements of the resistance of grain boundaries in Y$_{1-x}$Ca$_x$Ba$_2$Cu$_3$O$_{7-d}$ between the superconducting transition temperature, $T_c$, and room temperature. Below $T_c$ the normal state properties were assessed by applying sufficiently high currents. The behaviour of the grain boundary resistance *versus* temperature and of the conductance *versus* voltage are discussed in the framework charge transport through a tunnel barrier. The influence of misorientation angle, oxygen content, and calcium doping on the normal state properties is related to changes of the height and shape of the grain boundary potential barrier.

I. INTRODUCTION

Grain boundaries in high-$T_c$ cuprates have become a subject of intensive study after it became clear that these defects are the cause of severe reduction of the attainable critical currents in polycrystalline materials.[1-4] Elucidating the electronic properties of grain boundaries is of great importance for the development of recipes to enhance their transport properties for the purpose of multigrain applications, like coated conductors. A widely discussed framework to describe the exponential decrease of critical current density and normal state conductivity with misorientation angle is the so-called model of band bending, as proposed by Mannhart *et al.*[5,6] Band bending at the grain boundary is expected to occur as a result of work-function differences, charging, and Fermi-level pinning at interface states, and leads to hole-depleted layers in the superconducting electrodes next to the boundary. According to this approach, the boundary is considered as consisting of three adjacent layers. The middle layer is the structurally distorted interface region. It is abutted by two charge depleted layers in the otherwise undisturbed crystals, which width is controlled by the electrostatic screening length. The thickness of the depletion layers, $l$, has been proposed to be related to the grain boundary potential, $V_i$, by $l \sim (2\varepsilon_0\varepsilon_r V_i/en)^{1/2}$, where $\varepsilon_r$ is the dielectric constant perpendicular to the grain boundary interface, and $n$ the density of mobile charge carriers in undisturbed material. As not all of these parameters have been accurately determined, it has not been possible to quantitatively determine the electronic width of the boundary with this approach. Only very recently the potential drop and spatial extent of the grain boundary potential barrier in $Y_{1-x}Ca_xBa_2Cu_3O_{7-d}$ have been measured using electron holography in a transmission electron microscope.[7]

Calcium doping, both globally and locally, has been shown to increase the transport properties of both low and high angle grain boundaries in $Y_{1-x}Ca_xBa_2Cu_3O_{7-d}$.[8-14] Schmehl et al. observed over an order of magnitude enhancement in the critical current density and normal state conductivity of 24° symmetric [001] tilt grain boundaries when the films were homogeneously doped.[15] This discovery was met with significant enthusiasm, but the mechanisms by which calcium doping improves the grain boundary properties are not established. The $Ca^{2+}$ ions substitute for $Y^{3+}$ in the blocking layers between the Cu-O planes, leading to a creation of additional holes in the Cu-O planes. If this occurs preferentially in the grain boundary, calcium is speculated to reduce the positive electrostatic barrier potential in height. This leads to a decrease of the electrostatic screening length and thus a reduction of the spatial extent of the depletion zones enveloping the grain boundary. In addition, calcium is suggested to increase the local carrier density in the hole depleted zones, which independently decreases the spatial extent of band bending at the grain boundary.

In this study we have focussed on the normal state properties of grain boundaries in $Y_{1-x}Ca_xBa_2Cu_3O_{7-d}$. The properties of grain boundaries below the superconducting transition temperature have been investigated intensively. However, only few data are available for higher temperatures[16-18] due to the complications arising from the high resistance $Y_{1-x}Ca_xBa_2Cu_3O_{7-d}$ in its normal state. We have investigated grain boundaries up to room temperature in order to enhance the understanding of the electronic structure of the grain boundary.

II. EXPERIMENTAL PROCEDURE

Epitaxial thin films of $Y_{1-x}Ca_xBa_2Cu_3O_{7-d}$ were grown using pulsed laser deposition from sintered targets with compositions $x = 0$ and 0.2. The films were deposited onto $10 \times 5$ mm² bicrystalline $SrTiO_3$ subtrates with grain boundary misorientation angles of 18°, 24°, 30°, 37° and 45°. All $SrTiO_3$ bicrystals contained symmetric [001] grain boundaries, except for those with a misorientation angle of 45°, which were asymmetric. Some properties of the films are given in Table I. Tracks of 2 – 4µm wide were patterned by standard UV photolithography and argon ion beam milling. Current biased transport measurements up to 100 mA were carried out between 5 and 300 K using a cryogenic dip probe.

In order to measure the resistance of a grain boundary of the material in its normal state, Wheatstone bridges were patterned over the grain boundaries. The resistance of a grain boundary is significantly less than the resistance of a reasonable length of adjoining tracks, which makes it hard to separate the contributions of the grain boundary and the track to the resistance.[18] For this reason, the Wheatstone bridge measurement technique was developed, originally for grain boundaries in CMR materials[19] and later for grain boundaries in $YBa_2Cu_3O_{7-d}$.[16-18] If the bridge structure is aligned with a grain boundary, the symmetry of the geometry ensures that all the resistance contributions balance to zero except for those arising from the grain boundary. In order to increase accuracy of the method, a Wheatstone bridge geometry that straddles the grain boundary 20 times (per arm) was applied. The resistance of the grain boundary is proportional to the ratio of the measured potential difference and the applied current.

## III. RESULTS

Figure 1a shows the normal state resistance versus temperature, $R(T)$, characteristic for $YBa_2Cu_3O_{7-d}$ bicrystalline samples with misorientation angles between 18° and 45°. Above the superconducting transition temperature, $I(V)$ characteristics were linear in the assessed current range of 10 µA. Below the transition temperature, however, the Josephson current causes anomalous behaviour of the $I(V)$ characteristic at low voltages, and the normal state resistance is less well defined. Therefore, $I(V)$ characteristics were measured at currents up to 100 mA. The resistance was extracted in a voltage range sufficiently large to prevent the resistance value being affected by the Josephson current and associated phenomena, and small enough to exclude the influence of non-linear $I(V)$ behaviour at high voltages. Generally, this voltage was found to be around 10 mV.

Each $R(T)$ characteristic is obtained by taking the mean of 5 bridge devices on a single grain boundary. The results from a particular sample are typically very consistent from device to device. There is a 20 – 30% variation in the absolute magnitude, but the functional form of the resistance is very similar. Only the 24° grain boundaries showed a larger variation due to the small absolute magnitude of its resistance. Such variations are not surprising, since the boundaries themselves are extremely inhomogeneous and previous work on bicrystalline junction arrays has shown that macroscopic transport properties can vary greatly along the length of a single grain boundary.[20-23] There is also a small device to device error that is due to resistance imbalance of the bridge. These errors have been minimised by taking the mean of all the 5 devices on a single boundary.

The normal state resistance is obtained via different methods below and above the transition temperature, which causes a discontinuity of the *R*(*T*) characteristic at $T_c$. In some cases, a typical peak structure was observed at $T_c$. This is an intrinsic feature of Wheatstone bridges with an overall resistance much higher than the object that is measured with it, in this case the grain boundary. A large additional imbalance can be introduced in the bridge if the current is unequally distributed due to additional factors, e.g. due to small differences in $T_c$ in different arms of the bridge. Some data points close to $T_c$ were removed in order to enhance clarity. In addition, the grain boundary resistance has not always the same magnitude below and above $T_c$. This is likely to be caused by a small additional imbalance of the Wheatstone bridge due to variations in film thickness, imperfect lithography, etc. Below $T_c$ the (crystalline) $Y_{1-x}Ca_xBa_2Cu_3O_{7-d}$ is superconducting and any additional imbalance is lost. Figure 1b shows the resistance of grain boundaries with misorientation angles between 24° and 45° in $Y_{0.8}Ca_{0.2}Ba_2Cu_3O_{7-d}$. The presence of calcium in the films leads to significant reductions of the grain boundary resistance in comparison with $YBa_2Cu_3O_{7-d}$. The resistance for all four misorientation angles is at least a factor 2 lower. The grain boundary angles 24°, 30° and 37° have a resistance area products lower than $2\times10^{-12}$ ? $m^2$ at all temperatures between 5 K and 300 K. The resistance of the 37° boundary is even lower than that of the 30° boundary, which is likely to be caused by differences in grain boundary structure that differ from sample to sample. The values for the resistance area products are comparable with those reported by the Augsburg group for grain boundaries in $Y_{1-x}Ca_xBa_2Cu_3O_{7-d}$[5,6,12].

The insets of Figures 1a and 1b show the grain boundary resistance normalised at *T* = 300 K. Specifically the grain boundaries in $YBa_2Cu_3O_{7-d}$ exhibit a transition from almost temperature-independent behaviour for low(er) misorientation angles to a strongly activated behaviour at large misorientation. For the 24° and 30° samples the resistance at 100 K is 30 – 60 % higher than at 300 K, whereas for the 37 and 45° samples this is 140 – 220 %. Although this trend is less strong in $Y_{0.8}Ca_{0.2}Ba_2Cu_3O_{7-d}$ (insert Figure 1b), also these data show a clear correlation between the misorientation angle and the temperature dependence of the resistance of a grain boundary.

The normal state resistance for 30° grain boundaries in $YBa_2Cu_3O_{7-d}$ and $Y_{0.8}Ca_{0.2}Ba_2Cu_3O_{7-d}$ is shown in Figure 2. The strong reduction of the grain boundary resistance due to the calcium doping is clearly illustrated in Figure 2a. The influence of oxygen content is shown for both samples with and without calcium. A $YBa_2Cu_3O_{7-d}$ sample was deoxygenated by annealing for 7 hours at 500°C in 0.002 atm. of $O_2$, which resulted in a decrease of $T_c$ to 41 K. A $Y_{0.8}Ca_{0.2}Ba_2Cu_3O_{7-d}$ sample was annealed after deposition in a low oxygen pressure (15 mbar), which resulted in a $T_c$ of 76 K. In both cases, the lower oxygen content in the film leads to a higher resistance of a 30° grain boundary. In order to elucidate the temperature dependence of the resistance characteristics in Figure 2a, all these data are shown normalised at 300 K in Figure 2b. The lower oxygen content leads clearly to an increased temperature dependence of the resistance in both the plain and the calcium doped film. Nevertheless, the temperature dependence of a fully oxygenated 37° boundary is larger, in spite of the smaller absolute value of its resistance.

Current density – voltage, $J(V)$, curves were measured with single tracks and Wheatstone bridges at a range of temperatures between 5 K and the transition temperature. The behaviour of the $J(V)$ curves at low temperatures is reasonably consistent for the different devices on one sample. Differences between measurements from different devices can be caused by variations of grain boundary properties along its length. The functional form of the $J(V)$ curves, like those of the $R(T)$ curves, is consistent from device to device. Figure 3 shows $J(V)$ curves at T = 6 K of grain boundaries in $YBa_2Cu_3O_{7-\delta}$ and $Y_{0.8}Ca_{0.2}Ba_2Cu_3O_{7-d}$ for four different misorientation angles. The $J(V)$ characteristics of the higher angles can be measured over a much greater voltage range, because of their higher resistance. The $J(V)$ curves show a distinct non-linear behaviour, specifically the grain boundaries with higher misorientation angles. Non-linear $J(V)$ curves were also observed in measurements on single tracks across the grain boundary, and previously in both optimally doped[18], and underdoped[24] $YBa_2Cu_3O_{7-d}$ samples.

In order to assess the non-linearity of the $J(V)$ curves, conductivity *versus* voltage ($G(V) = \delta J / \delta V$) characteristics have been extracted from the measured $J(V)$ curves. Figure 4 compares $G(V)$ characteristics at $T$ = 6 K for grain boundaries with misorientation angles of 30°, 37° and 45° in $YBa_2Cu_3O_{7-d}$ and $Y_{0.8}Ca_{0.2}Ba_2Cu_3O_{7-d}$. Due to the lower resistance of 24° grain boundaries, these could only be measured in a smaller voltage range, which means that little information can be obtained from their $G(V)$ characteristics. Therefore, data from 24° grain boundaries have been omitted from Figure 4. Generally, the conductance increases with increasing voltage up to a point ($V,G$) where the current density causes local heating. A local increase of the temperature transforms parts of the grain boundary to a state with a higher resistance. In these measurements, local heating sets in at a power ($P = GV^2$) between 0.1 and 1 µW/µm². Therefore, the voltage range over which the conductance can be assessed depends on the magnitude of the conductance.

Analysis of $J(V)$ characteristics between 6 K and the superconducting transition temperature reveals that the normal state properties of the grain boundaries are essentially temperature independent between the gap voltage (10 - 20 mV) and the point where heating sets in. Figure 5 shows $G(V)$ characteristics for a range of temperatures of grain boundaries with a misorientation angle of 45° in YBa$_2$Cu$_3$O$_{7-d}$ and Y$_{0.8}$Ca$_{0.2}$Ba$_2$Cu$_3$O$_{7-d}$. Due to the high conductivity of grain boundaries in Y$_{0.8}$Ca$_{0.2}$Ba$_2$Cu$_3$O$_{7-d}$, the voltage range over which the conductance can be assessed is smaller than in the case of YBa$_2$Cu$_3$O$_{7-d}$. Local heating causes that the assessable voltage range decreases with increasing temperature.

## IV. DISCUSSION

The magnitude and functional form of the grain boundary resistance as a function of temperature were compared with different models for charge transport over a barrier.[17] It was found that the behaviour is inconsistent with charge transport due to thermal emission or variable range hopping. Instead, it is more characteristic of tunnelling through a potential barrier as a result of charging and band bending at the grain boundary. Following this conclusion, the grain boundary has been modelled as a tunnel barrier with a trapezoidal shape. The model predicts absolute values of the transport current by taking into account the Fermi velocity of the charge carriers, the Fermi occupation function for the relevant band structure and the tunnelling probability according to the WKB approximation. The model has only two free parameters: the height of the potential barrier and the width of the adjacent region over which band bending extends. Values for the barrier height and width were found to be in the order of 0.2 eV (above the Fermi energy) and 1 – 3 nm, respectively.

Within the framework of this model, the barrier height is associated with the magnitude of the transport current, whereas the trapezoidal shape is linked with its temperature dependence. The temperature dependence can be explained by taking into account that, with increasing temperature, charge carriers will have a higher average energy, and will, due to the trapezoidal potential barrier, encounter a narrower effective tunnel barrier. According to this approach, a rectangular potential barrier should lead to a temperature independent resistance as a function of temperature. This implies that the 24° and 30° boundaries have nearly rectangular barrier due to the almost temperature independent resistance, as shown in the insets of Figure 1. The 37° and 45° boundaries have a temperature dependence, which is up to 5 times higher. Thus, this suggests a transition between 30° and 37° from a very rectangular potential barrier to a trapezoidal barrier that extends further in the regions adjacent to the grain boundary. According to the band-bending model this is associated with charge-depleted regions, but it can also have microstructural causes as changes in stoichiometry, material density, bonding, lattice strain, structural disorder, etc.

The presence of a trapezoidal barrier at the grain boundary is expected to also influence the shape of its *J(V)* and *G(V)* characteristics. With increasing voltage charge carriers will have a higher average energy and will encounter a narrower effective tunnel barrier, which leads to non-linearity of *J(V)* characteristics and a positive correlation between voltage and conductance. In the case of a rectangular potential barrier, however, a more linear *J(V)* characteristic is expected, which amounts to a reduced dependence of the conductance on the voltage. Due to the early onset of heating in high conductivity barriers, only a limited voltage range can be assessed in the case of the grain boundaries for which a rectangular barrier is expected, those with misorientation angles of 24° and 30°. The *G(V)* characteristic of 24° boundaries is entirely dominated by the Josephson current and the associated RSJ behaviour. However, in the case of the 30° boundary in $YBa_2Cu_3O_{7-d}$, the correlation between conductance and voltage is limited above the voltage where RSJ behaviour plays a role, above approximately 10 mV (Figure 4). This is in agreement with the observation that the resistance of a 30° boundary is almost independent of temperature (Figure 1a). For the lower conductivity 37° and 45° boundaries in $YBa_2Cu_3O_{7-d}$, and 45° boundary $Y_{0.8}Ca_{0.2}Ba_2Cu_3O_{7-d}$, a large voltage range can be assessed, and the presence of a trapezoidal barrier is confirmed by the strong positive correlation between conductance and voltage.

The band bending model predicts that if the carrier density is reduced, the depletion length increases, and the band structure is distorted further into the material. Figure 2a shows that a lower oxygen content leads to an increase of the resistance of grain boundaries in $YBa_2Cu_3O_{7-d}$ and $Y_{0.8}Ca_{0.2}Ba_2Cu_3O_{7-d}$. Moreover, the increased temperature dependence of the resistance, as shown in Figure 2b, suggests a change of the potential barrier to a more trapezoidal shape. As this is associated with an increased spatial extent of band bending, it indicates that a lower oxygen content leads to a reduced mobile carrier density at the grain boundary.

The presence of calcium is proposed to influence the height and width of the grain boundary potential barrier via different mechanisms, as elucidated in the introduction. The preferential replacement of $Y^{3+}$ by $Ca^{2+}$ is speculated to reduce the positive electrostatic potential of the grain boundary in height. The electrostatic screening length is accordingly decreased with the root of the grain boundary potential. Independently, calcium is suggested to increase the local carrier density in the hole depleted zones adjoining the grain boundary, thus reducing principally the width of the barrier. The lower resistance of grain boundaries in $Y_{0.8}Ca_{0.2}Ba_2Cu_3O_{7-d}$ in comparison with those in $YBa_2Cu_3O_{7-d}$ (Figures 1 and 2) indicates a reduced height of the grain boundary potential barrier. In addition, the reduced temperature dependence of the resistance (Figures 1 insets and 2b) suggests a change in the shape of the barrier from trapezoidal to more rectangular due to the presence of calcium. This seems to be specifically the case for the grain boundaries for which a more trapezoidal barrier is expected, namely those with misorientation angles of 37° and 45°. The *G(V)* characteristics of 45° boundaries (Figure 5) confirm this. The conductivity of a 45° boundary in $YBa_2Cu_3O_{7-d}$ increases with a factor 3 over a voltage range of 80 mV, whereas in $Y_{0.8}Ca_{0.2}Ba_2Cu_3O_{7-d}$ this is a factor 1.4 over a voltage range of 40 mV. We will try to relate these observations to band bending and the suggested role(s) of calcium in this model. As the depletion length decreases with the root of the barrier potential ($l \propto V_i^{1/2}$), a decrease of the height of the barrier potential will lead to a lower and more trapezoidal barrier. Therefore, a reduction of the electrostatic potential due to the preferential substitution of $Y^{3+}$ by $Ca^{2+}$ can explain the lower normal state resistance of calcium doped films, but can not explain the decreased temperature and voltage dependence of the *R(T)* and *G(V)* characteristics, respectively. However, an increase of the local carrier density in the charge depleted zones due to calcium doping will mainly affect the spatial extent of the grain boundary, and thus lead to a change of the shape of the potential barrier to more rectangular. Therefore, the observed changes of the height and the shape of the potential barrier can only be explained by taking into account the effect of calcium on both the electrostatic potential and the mobile carrier density.

## V. CONCLUSION

The Wheatstone bridge geometry was found to be an accurate and reproducible technique to measure the normal state properties of grain boundaries in $Y_{1-x}Ca_xBa_2Cu_3O_{7-d}$ above $T_c$. The decreasing resistance with temperature was explained by considering the shape of the potential barrier at the grain boundary. Grain boundaries with higher misorientation angles were proposed to exhibit a potential barrier with a more trapezoidal shape. This was confirmed by the positive correlation between differential conductance and voltage. The oxygen content of the $Y_{1-x}Ca_xBa_2Cu_3O_{7-d}$ is proposed to have a large influence on the spatial extent of the grain boundary, thus confirming earlier assumptions that the oxygen content is strongly linked to charge depletion at the grain boundary. The presence of calcium results in a significant enhancement of the normal state properties at all temperatures between 5 K and 300 K. Also the temperature dependence of the resistance and the voltage dependence of the conductance decrease, which suggests that calcium doping changes the shape of the grain boundary potential barrier from trapezoidal to more rectangular. This cannot be explained by a reduction of the electrostatic barrier potential, but only by taking into account the effect of calcium on both the electrostatic potential and the mobile carrier density.

FIGURE CAPTIONS

Table I. Thickness, resistivity at 300 K and superconducting transition temperature of the $Y_{1-x}Ca_xBa_2Cu_3O_{7-d}$ films used for this study.

Figure 1 a) Resistance area products for grain boundaries in $YBa_2Cu_3O_{7-d}$. The misorientation angles of the grain boundaries range, as indicated, from 18° to 45°. The inset shows the resistance area products for the four highest misorientation angles normalised at 300 K. b) Resistance area products for grain boundaries with misorientation angles between 24° and 45° in $Y_{0.8}Ca_{0.2}Ba_2Cu_3O_{7-d}$. The inset shows the resistance area products normalised at 300 K.

Figure 2. a) Resistance area products for 30° grain boundaries in $YBa_2Cu_3O_{7-d}$ and in $Y_{0.8}Ca_{0.2}Ba_2Cu_3O_{7-d}$. Results for samples with reduced oxygen content are also shown. The arrows indicate the change of the resistance *versus* temperature characteristic due to a decreased oxygen content. The behaviour of a grain boundary with a misorientation angle of 37° in $YBa_2Cu_3O_{7-d}$ is shown for comparison. b) All resistance *versus* temperature characteristics of Figure 2a normalised at 300 K. The arrows indicate the change of the temperature dependence of the resistance due to a decreased oxygen content.

Figure 3. a) Current voltage characteristics at 6 K of grain boundaries in $YBa_2Cu_3O_{7-d}$ with misorientation angles between 24° and 45°. B) Current voltage characteristics of grain boundaries in $Y_{0.8}Ca_{0.2}Ba_2Cu_3O_{7-d}$. Note that the x- and y-axis scales are identical in a) and b).

Figure 4. Differential conductivity ($G = \delta J / \delta V$) at 6 K as a function of voltage for grain boundaries with three different misorientation angles in $YBa_2Cu_3O_{7-d}$ and $Y_{0.8}Ca_{0.2}Ba_2Cu_3O_{7-d}$ films. These $G(V)$ characteristics were extracted from the $J(V)$ characteristics in Figure 3. Note that the scale at the y-axis is logarithmic. Some data points close to zero voltage are removed because of the Josephson current peak. The proposed difference in shape of the potential barrier is qualitatively represented for grain boundaries with misorientation angles of 30° and 37° / 45°.

Figure 5. Differential conductivity ($G = \delta J / \delta V$) as a function of voltage for grain boundaries with a misorientation angle of 45° in a) $YBa_2Cu_3O_{7-d}$ and b) $Y_{0.8}Ca_{0.2}Ba_2Cu_3O_{7-d}$ films. Conductivity data are shown for a range of temperatures with intervals of 6 K between 6 K and a) 84 K, b) 66K. Note that the (linear) x- and y-axis scales are not identical in a) and b). Noise features have been removed from these data using a fast Fourier transform smoothing routine.

**Table I.**

| $q$ | $x$ | d (nm) | $r$ ($\Omega$m) | $T_c$ (K) |
|---|---|---|---|---|
| 18° | 0 | 187 | 9.0 | 89.5 |
| 24° | 0 | 100 | 11.0 | 88.5 |
| 30° | 0 | 210 | 4.4 | 92.1 |
| 37° | 0 | 77 | 10.0 | 81.7 |
| 45° | 0 | 135 | 3.2 | 89.2 |
| 24° | 0.2 | 230 | 5.2 | 81.0 |
| 30° | 0.2 | 225 | 7.7 | 82.3 |
| 37° | 0.2 | 200 | 7.1 | 81.0 |
| 45° | 0.2 | 210 | 10.7 | 83.4 |

**Figure 1.**

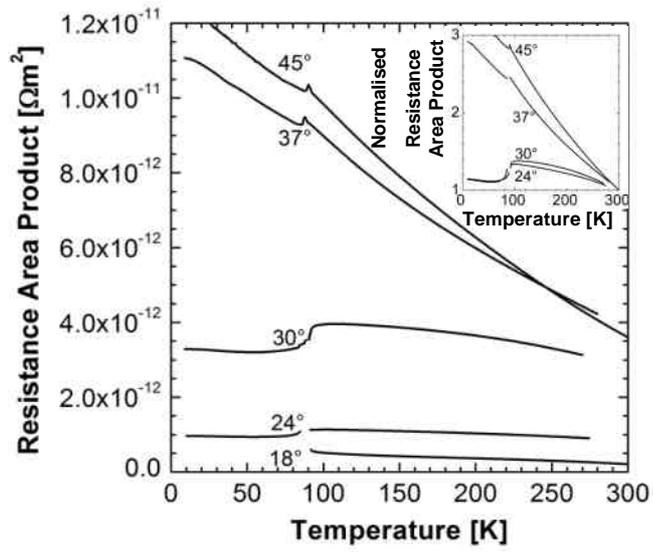

a

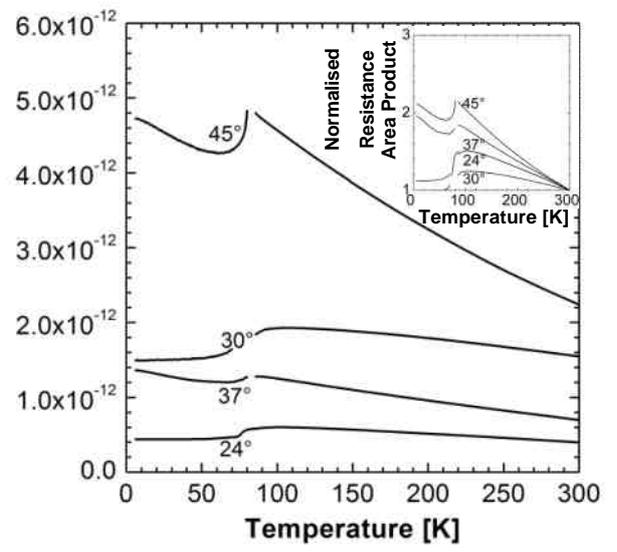

b

**Figure 2.**

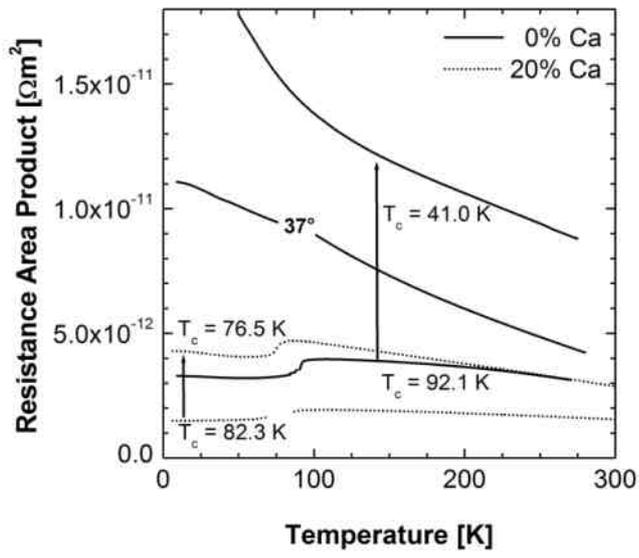 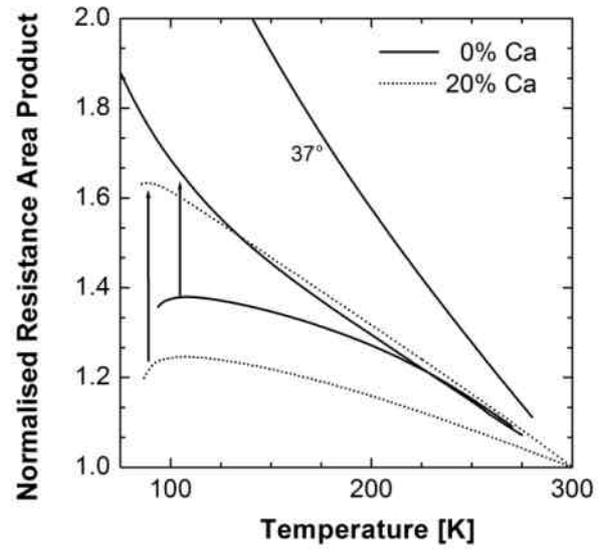

a　　　　　　　　　　　　　　　　　　b

**Figure 3.**

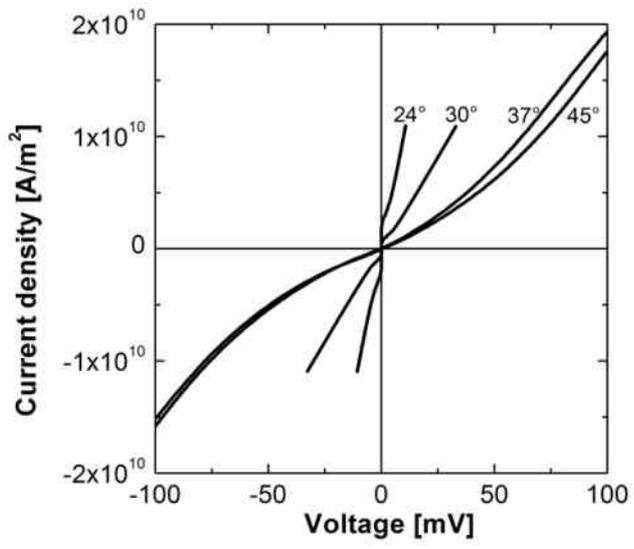 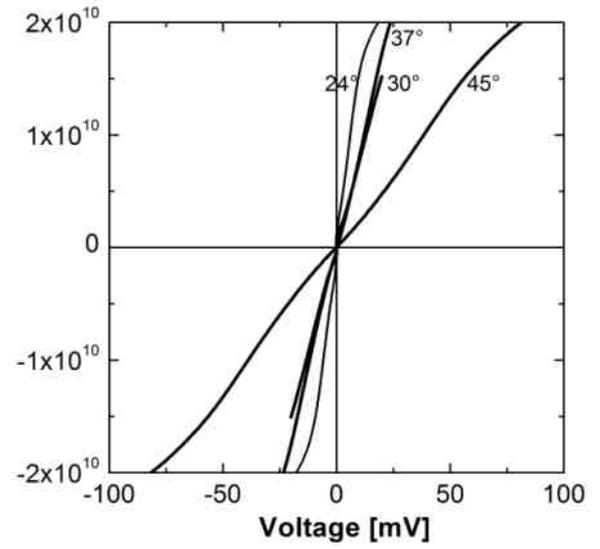

a b

**Figure 4.**

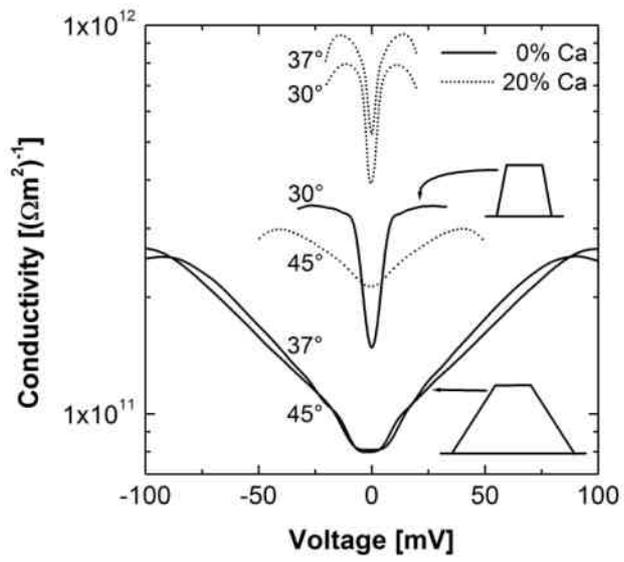

**Figure 5.**

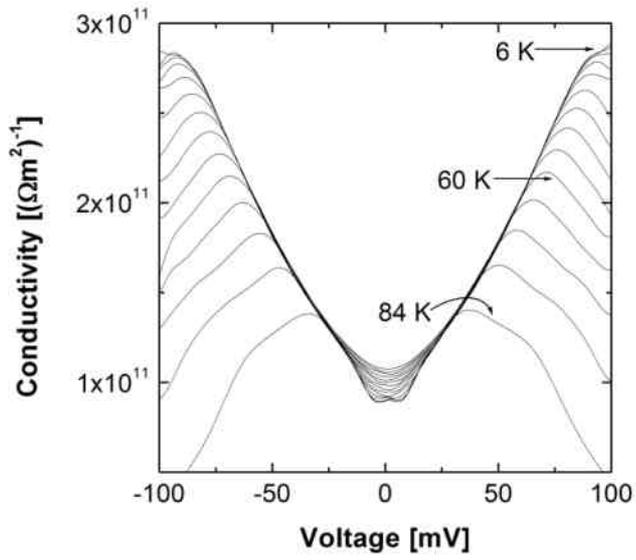
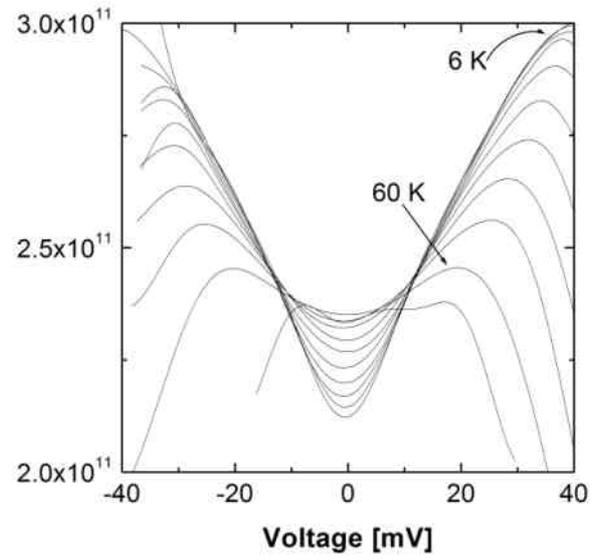

a b